\documentclass[twocolumn,amsmath,amssymb]{revtex4}
\usepackage{dcolumn}
\usepackage{bm}
\begin{document}

\noindent \textbf{Barbosa et al. Reply to ``Comment on 'Secure Communication using mesoscopic coherent states', Barbosa et al, Phys Rev Lett 90, 227901", Z.L. Yuan, A.J. Shields, Phys. Rev. Lett. 94, 048901(2005) }\\
\indent The authors of Ref.~\cite{yuan05} claim that the
$\alpha\eta$ protocol \cite{barbosa03} is entirely equivalent to a
classical stream cipher utilizing no quantum phenomena, a common
misconception.  Indeed, a paper has already appeared in Phys. Lett.
A \cite{nishioka04} to a similar effect, and we have responded
\cite{yuen04_2}.  Still, we welcome the opportunity to clarify the
situation for a wider audience.

Before proceeding, we need to point out a terminological shortcoming in Ref.~\cite{yuan05} that further confuses the issue.  The term \emph{one-time pad} is consistently used in Ref.~\cite{yuan05} to mean a conventional
(classical) cipher with a shared secret key, and should always be
interpreted as such in reading Ref \cite{yuan05}.  Furthermore, one cannot state that such a classical cipher with a shared secret key is ``not secure", as security needs to be quantified.  The main points of
Ref.~\cite{yuan05} are that a) the security of $\alpha\eta$ is ``unrelated to
quantum noise'' for direct encryption, and b) $\alpha\eta$ cannot be used to
``expand the secret information they share'' for key generation.

Our response, in sum, is that there is a distinction between
what has been achieved in our experiments and what is in principle
possible with the $\alpha\eta$ protocol.  It is true that our
experiments, thus far, have not operated in a regime which allows key generation and we have not
claimed that they have; but in principle key generation is possible as already
indicated in Ref.~\cite{barbosa03}, a crucial point missed in \cite{yuan05}.  For
direct encryption our experiments do utilize quantum noise in an
essential way that could be done classically in principle,
though not in practice.  In principle, $\alpha\eta$ may perform
far beyond a conventional cipher, especially with added techniques
and lower signal energy.  These are described in some detail
in Ref.~\cite{yuen04} and briefly discussed in Ref.~\cite{yuen04_2}.

The mistake in claiming that $\alpha\eta$ cannot generate new
key derives from ignoring the optimal quantum receiver performance
between Eve who does not know the key $K$ when she makes her quantum
measurement, and Bob who does.  It follows from quantum detection
theory \cite{helstrom76} that the bit-error rates in discriminating an antipodal
signal set with the optimum quantum-receiver, an optimum quantum phase-measurement, and balanced heterodyne-detection are
\begin{equation}\label{error}
P_e^{\mathrm{opt}}\sim e^{-4S},\;\;
P_e^{\mathrm{ph}}\sim e^{-2S},\;\;
P_e^{\mathrm{het}}\sim e^{-S},
\end{equation}
respectively, where $S$ is the signal photon number.  Note that while the transmitted signal set is M-ry, use of the secret key collapses the signal set to binary antipodal \cite{barbosa03}.  Further note that the optimum quantum-receiver performance can \emph{only} be achieved if the secret key is used during measurement; something only Bob can do.  This is a quantum effect with \emph{no} classical analog, and represents a key generation principle fundamentally different from the well known BB84 and Ekert protocols.

For bounding Eve's performance, we may grant her a full copy of the
quantum signal and allow her to perform any quantum measurement.  We
have shown numerically that Eve's optimum quantum receiver's
performance approaches that of the optimal quantum phase-measurement
-- still a factor of two in the exponent worse than Bob's optimum
receiver. As an example with a mesoscopic signal level $S\sim7$, one
has $P_e^{\mathrm {opt}}\sim 10^{-12}$, $P_e^{\mathrm{ph}}\sim
10^{-6}$, and $P_e^{\mathrm{het}}\sim 10^{-3}$.  At a 1 Gbps data
rate, Alice/Bob are likely to generate $\sim 10^3$ or $\sim 10^6$
new key bits per second (with privacy amplification) depending on
Eve's exact measurement.  Note that while Eqn.~\ref{error} also
applies for large $S$, the resulting key generation rate would be
extremely low.

For direct encryption, the coherent-state quantum noise is
used in the experiments of Ref.~\cite{barbosa03} to provide high speed
randomization that can in principle, but not now in practice, be
done classically.  This randomization is what distinguishes
$\alpha\eta$ from a conventional cipher.  Moreover, with the additional mechanisms like deliberate state-randomization, as described briefly in Ref.~\cite{yuen04}, information-theoretic security against known-plaintext attacks may be obtained -- something that has yet to be shown for any conventional cipher.  Addressing another criticism in Ref.~\cite{yuan05}, using
many bits of the running key does not reduce the efficiency of the
seed key $K$.  The efficiency and security of conventional and
quantum cryptographic systems are much more subtle issues than the
treatment available in the literature may suggest. Some detailed
discussions of these issues, particularly those related to
$\alpha\eta$, are given in Ref.~\cite{yuen04}.\\\\

\noindent Horace Yuen, Eric Corndorf, Geraldo Barbosa, and Prem Kumar\\
\indent\small{Center for Photonic Communication and Computing\\\indent Northwestern University\\\indent 2145 Sheridan Road, Evanston, IL 60208\\\indent yuen@ece.northwestern.edu}


\begin{thebibliography}{6}
\expandafter\ifx\csname natexlab\endcsname\relax\def\natexlab#1{#1}\fi
\expandafter\ifx\csname bibnamefont\endcsname\relax
  \def\bibnamefont#1{#1}\fi
\expandafter\ifx\csname bibfnamefont\endcsname\relax
  \def\bibfnamefont#1{#1}\fi
\expandafter\ifx\csname citenamefont\endcsname\relax
  \def\citenamefont#1{#1}\fi
\expandafter\ifx\csname url\endcsname\relax
  \def\url#1{\texttt{#1}}\fi
\expandafter\ifx\csname urlprefix\endcsname\relax\def\urlprefix{URL }\fi
\providecommand{\bibinfo}[2]{#2}
\providecommand{\eprint}[2][]{\url{#2}}

\bibitem[{\citenamefont{Yuan and Shields}(2005)}]{yuan05}
\bibinfo{author}{\bibfnamefont{Z.}~\bibnamefont{Yuan}} \bibnamefont{and}
  \bibinfo{author}{\bibfnamefont{A.}~\bibnamefont{Shields}},
  \bibinfo{journal}{Physics Review Letters} \textbf{\bibinfo{volume}{94}},
  \bibinfo{pages}{048901} (\bibinfo{year}{2005}).

\bibitem[{\citenamefont{Barbosa et~al.}(2003)\citenamefont{Barbosa, Corndorf,
  Kumar, and Yuen}}]{barbosa03}
\bibinfo{author}{\bibfnamefont{G.}~\bibnamefont{Barbosa}},
  \bibinfo{author}{\bibfnamefont{E.}~\bibnamefont{Corndorf}},
  \bibinfo{author}{\bibfnamefont{P.}~\bibnamefont{Kumar}}, \bibnamefont{and}
  \bibinfo{author}{\bibfnamefont{H.}~\bibnamefont{Yuen}},
  \bibinfo{journal}{Physics Review Letters} \textbf{\bibinfo{volume}{90}},
  \bibinfo{pages}{227901} (\bibinfo{year}{2003}).

\bibitem[{\citenamefont{Nishioka et~al.}(2004)\citenamefont{Nishioka, Hasegawa,
  Ishizuka, Imafuku, and Imai}}]{nishioka04}
\bibinfo{author}{\bibfnamefont{T.}~\bibnamefont{Nishioka}},
  \bibinfo{author}{\bibfnamefont{T.}~\bibnamefont{Hasegawa}},
  \bibinfo{author}{\bibfnamefont{H.}~\bibnamefont{Ishizuka}},
  \bibinfo{author}{\bibfnamefont{K.}~\bibnamefont{Imafuku}}, \bibnamefont{and}
  \bibinfo{author}{\bibfnamefont{H.}~\bibnamefont{Imai}},
  \bibinfo{journal}{Physics Letters A} \textbf{\bibinfo{volume}{327}},
  \bibinfo{pages}{28} (\bibinfo{year}{2004}).

\bibitem[{\citenamefont{Yuen et~al.}(2004)\citenamefont{Yuen, Kumar, and
  Corndorf}}]{yuen04_2}
\bibinfo{author}{\bibfnamefont{H.}~\bibnamefont{Yuen}},
  \bibinfo{author}{\bibfnamefont{P.}~\bibnamefont{Kumar}}, \bibnamefont{and}
  \bibinfo{author}{\bibfnamefont{E.}~\bibnamefont{Corndorf}},
  \bibinfo{author}{\bibfnamefont{R.}~\bibnamefont{Nair}},
  \bibinfo{journal}{quant-ph/0407067, submitted to Phy. Lett. A}
  (\bibinfo{year}{2004}).

\bibitem[{\citenamefont{Yuen}(2004)}]{yuen04}
\bibinfo{author}{\bibfnamefont{H.}~\bibnamefont{Yuen}},
  \bibinfo{journal}{quant-ph/0311061}  (\bibinfo{year}{2004}).

\bibitem[{\citenamefont{Helstrom}(1976)}]{helstrom76}
\bibinfo{author}{\bibfnamefont{C.}~\bibnamefont{Helstrom}},
  \emph{\bibinfo{title}{Quantum Detection and Estimation Theory}}
  (\bibinfo{publisher}{Academic}, \bibinfo{address}{New York},
  \bibinfo{year}{1976}).
\end{thebibliography}
\end{document}